\documentclass[runningheads,a4paper]{llncs}

\usepackage{amssymb}
\usepackage{graphicx}

\usepackage{amsmath}
\usepackage{tikz}
\usetikzlibrary{arrows,snakes,backgrounds}
\usetikzlibrary{automata,positioning}
\usepackage{calc}
\usepackage{ifthen}
\usepackage{multirow}
\usepackage{hhline}

\usepackage{arydshln}

\usepackage{graphicx}
\usepackage{caption}
\usepackage{subcaption}

\usepackage{algorithm,algpseudocode}
\usepackage{url}
\usepackage{multirow}
\usepackage{rotating}
\usepackage{tipa}

\usepackage{url}
\usepackage{soul} 
\urldef{\mailsa}\path|{alfred.hofmann, ursula.barth, ingrid.haas, frank.holzwarth,|
\urldef{\mailsb}\path|anna.kramer, leonie.kunz, christine.reiss, nicole.sator,|
\urldef{\mailsc}\path|erika.siebert-cole, peter.strasser, lncs}@springer.com|    
\newcommand{\keywords}[1]{\par\addvspace\baselineskip
\noindent\keywordname\enspace\ignorespaces#1}

\begin{document}

\mainmatter  

\title{SS4MCT: A Statistical Stemmer for Morphologically Complex Texts}

\titlerunning{SS4MCT: A Statistical Stemmer for Morphologically Complex Texts}

%
%
\author{Javid Dadashkarimi%
\and  Hossein Nasr Esfahani  \and Heshaam Faili \and Azadeh Shakery}

\urldef{\mailsa}\path|{dadashkarimi, h_nasr, shakery, hfaili}@ut.ac.ir|    

\institute{}
\institute{School of Electrical and Computer Engineering,\\
College of Engineering,\\
University of Tehran, Tehran, Iran. \\
\mailsa\\}
\authorrunning{Dadashkarimi et al.}

%
%

\toctitle{SS4MCT: A Statistical Stemmer for Morphologically Complex Texts}
\tocauthor{Authors' Instructions}
\maketitle

\begin{abstract}
There have been multiple attempts to resolve various inflection matching problems in information retrieval. Stemming is a common approach to this end. Among many techniques for stemming, statistical stemming has been shown to be effective in a number of languages, particularly highly inflected languages. In this paper we propose a method for finding affixes in different positions of a word.
Common statistical techniques heavily rely on string similarity in terms of prefix and suffix matching.
Since infixes are common in irregular/informal inflections in morphologically complex texts, it is required to find infixes for stemming.
In this paper we propose a method whose aim is to find statistical inflectional rules based on minimum edit distance table of word pairs and the likelihoods of the rules in a language.
These rules are used to statistically stem words and can be used in different text mining tasks. 
Experimental results on CLEF 2008 and CLEF 2009 English-Persian CLIR tasks indicate that the proposed method significantly outperforms all the baselines in terms of MAP.
\end{abstract}
\vspace{-0.2cm}
\section{Introduction}
\vspace{-0.3cm}

Uniforming different inflections of words is a required task in a wide range of text mining algorithms, including, but not limited to text classification, text clustering, document retrieval, and language modeling~\cite{Aggarwal:2012}. Stemming has been considered as a common approach for this goal in several studies~\cite{Krovetz:1993,Aggarwal:2012}.
Stemmers usually remove affixes from the words to present them in the form of their morphological roots. Conventional rule-based stemmers tailor the linguistic knowledge of experts. On the other hand, statistical stemmers  provide language-independent approaches which generally group related words based on various string-similarity measures. Such approaches often involve n-grams;
equivalence classes can be formed from words that share the same properties: word-initial letter n-grams, common n-grams throughout the word, or by refining these classes with clustering techniques. 
This kind of statistical stemming has been shown to be effective for many languages, including English, Turkish, and Malay. 
 For example, Bhat introduced a method for Kannada where the similarity of two words
is determined by three distance measures based on prefix and suffix matching and the first mismatch point in the words~\cite{Bhat:2013}.

Defining precise rules for morphologically complex texts, especially for the purpose of infix removal is sometimes impossible~\cite{Dadashkarimi:2014}.
Informal/irregular forms usually do not obey the conventional rules in the languages.
For instance, `\textit{$\hat{\text{kh}}$unh}' (home) is a frequent form for `\textit{$\hat{kh}$anh}' in Persian conversations or `\textit{goood}' and `\textit{good}' are used interchangeably in English tweets.

In this paper,  we propose a statistical technique for finding inflectional and derivation formations of words.
To this end, we introduce an unsupervised method to cluster all morphological variants of a word. 
The proposed algorithm learns linguistic patterns to match a word and its morphological variants based on a given large collection of documents, which is readily available on the Web.
A linguistic pattern captures a  transformation rule between a word and its morphological variant. The extracted rules indicate which letters in which positions of a word should be modified.
Affix characters, positions of the characters, operations on the characters based on the minimum edit distance (MED) algorithm (i.e., \textit{insertion} or \textit{deletion}) ~\cite{Levenshtein:1996}, and part-of-speech (POS) tag of the input word are the attributes of a rule. 
Our algorithm assigns a score to each rule,  indicating its confidence. The higher the frequency of a rule in the input collection, the higher the confidence value of that rule. Finally, a small subset of the obtained rules are selected based on their scores and a learned threshold as valid rules.
We demonstrate that using this subset for query expansion can significantly improve English-Persian CLIR performance compared to comprehensive baselines.

In Section~\ref{Formation Matching} we elaborate on the subject, in Section~\ref{Experiments} we assess its quality in an IR task, and in Section~\ref{Conclusion} we conclude the paper.

\vspace{-0.4cm}
\section{SS4MCT: A Statistical Stemmer for Morphologically Complex Texts}
\label{Formation Matching}
In this section we propose an unsupervised method for finding inflections in a language. To this end, we first introduce a transformation rule which is an edit path transforming the word $w$ into $w'$ based on  a number of actions. Our goal is to estimate the probability of each transformation rule $P(R)$ and compute the likelihood of generating inflections for a given term (i.e., $p(w'|w)$).
In Section \ref{Transformation Rules} and Section \ref{Probability of the Rules}  we introduce the proposed method in more details and in Section \ref{Evaluating the Algorithm} we propose an evaluation framework for the method.
\vspace{-0.3cm}
\subsection{Transformation Rules }
\label{Transformation Rules}
\vspace{-0.3cm}
Each transformation rule contains a number of actions transforming a word into an inflection. If two terms are $k$ points distant from each other, the rule that transforms the input term to the output term contains $k$ actions and the maximum likelihood POS tag of the input word. Each action consists of the following attributes: \textbf{c}, the character in difference, \textbf{p}, the position of that character (\textit{\textbf{b}egin}, \textit{\textbf{m}iddle}, and \textit{\textbf{e}nd}), and \textbf{o}, the corresponding operation on the character in the MED algorithm (\textit{deletion} or \textit{insertion}).
Intuitively we define a few general positions for affixes to prevent sparsity of the rules.
A \textit{substitution} operation can be replaced by a couple of \textit{insertion}/\textit{deletion} operations; therefore we ignore the \textit{substitution} operation. Table~\ref{table:affix} shows a number of examples for the rules.

\begin{table}[t]
	\centering
	\caption{Examples of transformation rules}
	\label{table:affix}
	\begin{tabular}{|c|c|c|c|c|c|c|c|c|}  \hline
		\multirow{2}{*}{$w$}                 &  \multirow{2}{*}{$\bar{w}$} & \multicolumn{7}{c|}{Transformation Rule} \\ \hhline{~~-------}
		& & o$_1$  & $p_1$ & $c_1$ & o$_2$ &$p_2$ & $c_2$ & POS     \\ \hline
		\textit{jhangrd} (tourist) & \textit{jhangrdi} (tourism)    & $i$ & $e$  & i  & - & -  & -  & \tiny{N\_SING} \\ \hline
		\textit{jhangrd}           & \textit{jhangrdan} (tourists)  & $i$ & $e$  & n  & $i$ & $e$  & a  & \tiny{N\_SING} \\ \hline
		\textit{jhangrd}           & \textit{jhangir} (proper noun) & $d$ & $m$  & i  & $i$& $e$  & d  & \tiny{N\_SING} \\ \hline
		\textit{ksart} (damage)    & \textit{ksarat} (damages)      & $i$ & $m$  & a  & - & - & -  & \tiny{N\_SING} \\ \hline
		\textit{shabe} (friend)    & \textit{ashab} (friends)       & $i$ & $b$  & a & $d$ & $e$  & e  & \tiny{N\_SING} \\ \hline
		\textit{jzirh} (island)    & \textit{jzair} (islands)       & $i$ & $m$  & a & $d$ & $e$  & e  & \tiny{N\_SING} \\ \hline
	\end{tabular}
\end{table}

\vspace{-0.5cm}
\subsection{Probability of the Rules}
\label{Probability of the Rules}
\vspace{-0.2cm}
To compute the probability of generating an inflection for a given term (i.e., $p(w'|w)$) we
can compute the transformation rule ($R$) between $w$ and $w'$ and estimate $p(w'|w)$ by $p(R)$.
To compute the probability of each rule we use a large collection of words extracted from a document collection. For each pair of words in the collection ($w$ and $w'$), we compute the rule for transforming $w$ into $w'$ and count the number of times this rule has happened in the collection. The higher the occurrences of a rule, the more likely it is to be a valid one.
Finally we estimate the rules probability with maximum likelihood estimator.

\vspace{-0.4cm}
\subsection{How to Evaluate the Algorithm}
\label{Evaluating the Algorithm}
\vspace{-0.2cm}
In this section we provide a framework for the stemming algorithm to evaluate its effectiveness. We use dictionary-based cross-lingual information retrieval (CLIR) to this end.
In highly inflected languages, bilingual dictionaries contain only original forms of the words. Therefore, in dictionary-based CLIR, retrieval systems are obliged either to stem documents and queries, or to leave them intact~\cite{Hashemi:2014,Dadashkarimi:2016,Rahimi:2015}, or expand the query with inflections. We opted the query expansion approach which is a widely used approach to compensate the shortage of inflections~\cite{Krovetz:1993,Cao:2008,Dadashkarimi:2014}. We used the following probabilistic framework to this end~\cite{Dadashkarimi:2014}:
\begin{align}
\label{dsiamb:javid:1}
p(c_{i,j}|q_i)=&\sum\limits_{i^\prime \neq i}^{}\Big(\sum\limits_{j^\prime=1}^{|\mathbf{c}_{i^\prime}|}p(c_{i,j},c_{i^\prime,j^\prime})+\sum\limits_{j^\prime=1}^{|\mathbf{c'}_{i^\prime}|}p(c_{i,j},c'_{i^\prime,j^\prime})\Big),\nonumber \\ 
p(c'_{i,j}|q_i)=&\sum\limits_{i\prime\neq i}^{}\sum\limits_{j\prime=1}^{|\mathbf{c}_{i\prime}|}p(c'_{i,j},c_{i\prime,j\prime}). 
\end{align}
where $q_i$ is a query term and $\textbf{c}_{i}$ is the set of translation candidates provided in a bilingual dictionary for $q_i$. $\textbf{c}'_{i}$ is the set of the most probable inflections of the words appeared in $\textbf{c}_{i}$ selected by a tuned threshold. Then, we compute the translation probability of $c_{i,j}$ or $c'_{i,j}$ for the given $q_i$. To avoid adding noisy terms, we only compute the joint probabilities between either a pair of translation candidates from the dictionary ($c_{i,j}$ and $c_{i^\prime,j^\prime}$) or a pair of a candidate from the dictionary and an inflection from the collection ($c_{i,j}$ and $c'_{i^\prime,j^\prime}$)~\cite{Dadashkarimi:2014}.

\begin{figure}[t]
	\centering
	\includegraphics[width=0.7\textwidth]{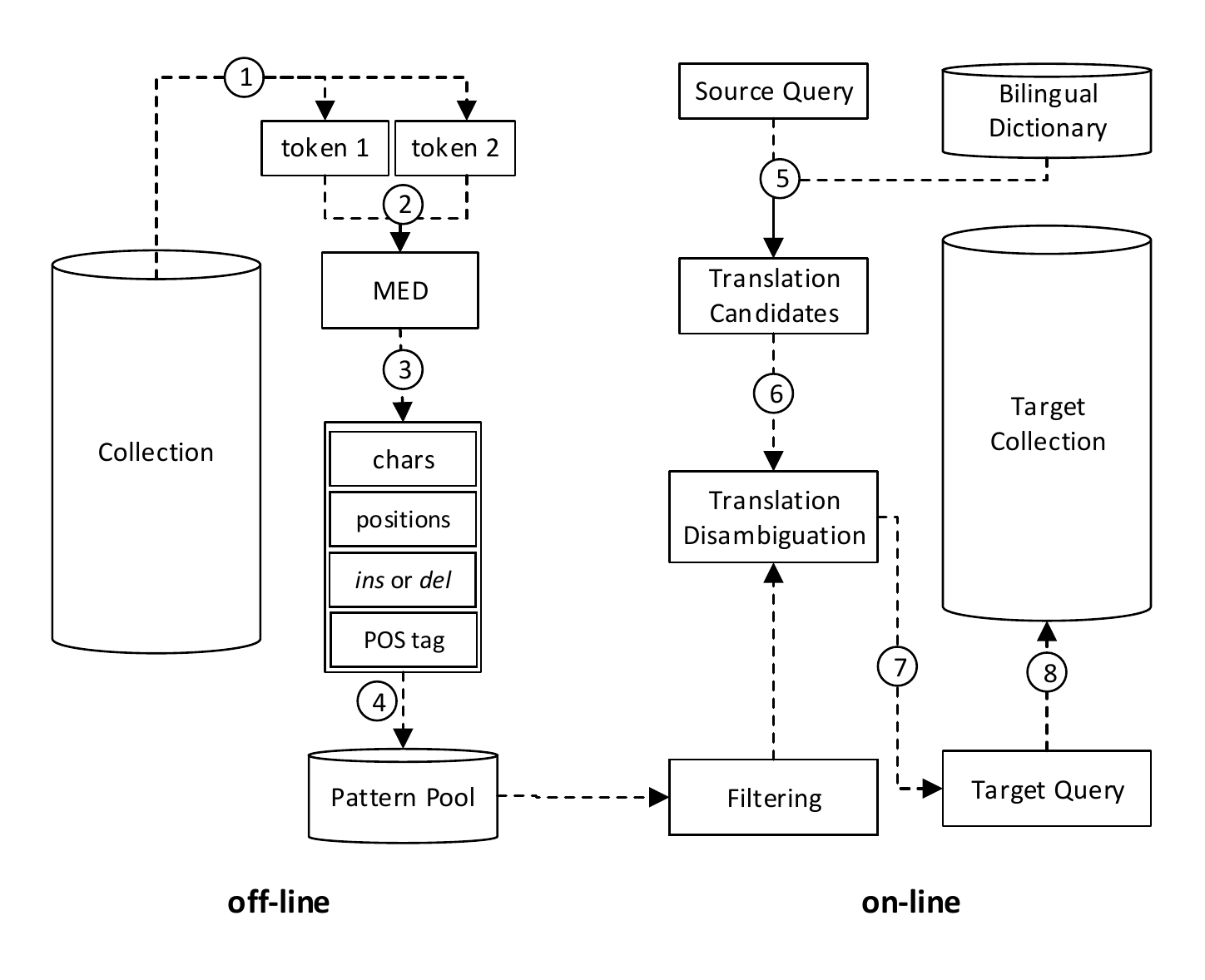}
	\caption{Outline the proposed SS4MCT and its evaluation framework.}
	\label{fig:pat and query expansion}
\end{figure}

Our goal is to find $\bar{\textbf{c}}_i$ using the proposed SS4MCT (i.e. set of top-ranked $\bar{c}_{i',j'}$ according to $p_t(\bar{c}_{i',j'}|c_{i,j})$) and then evaluate its impact on the performance of the CLIR task.
Figure~\ref{fig:pat and query expansion} shows the whole process of extracting rules (off-line part) and the evaluation framework (on-line part).



\vspace{-0.2cm}
\section{Experiments}
\label{Experiments}
\vspace{-0.2cm}

\subsection{Experimental Setup}
\vspace{-0.1cm}

The statistics of the collection used for both rule extraction and evaluation is provided in Table~\ref{tab:dataset}. We employed the statistical language modeling framework with Kullback-Leibler similarity measure of Lemur toolkit for our retrieval task. Dirichlet Prior is selected as our document smoothing strategy. Top 30 documents are used for the mixture pseudo-relevance feedback algorithm. Queries are expanded by the top 50 terms generated by the feedback model~\cite{Zhai:2001,Esfahani:2016}.
We removed Persian stop words from the queries and documents~\cite{Dadashkarimi:2016,Dadashkarimi:2014}.
We used STeP1~\cite{Shamsfard:2010} in our stemming process in Persian. We also stem the source English queries in all experiments with the Porter stemmer.
We use \textit{Google} English-Persian dictionary \footnote{https://translate.google.com/$\#$en/fa/} as the translation resource. Dadashkarimi et al., demonstrated that \textit{Google} has better coverage compared to other English-Persian dictionaries~\cite{Dadashkarimi:2014}. 
We have exploited 40 Persian POS tags in our experiments.\footnote{http://ece.ut.ac.ir/dbrg/bijankhan/}
The retrieval results are mainly evaluated by Mean Average Precision (MAP) over top 1000 retrieved documents. Significance tests are computed using two-tailed paired t-test with 95\% confidence. Precision at top 5 documents (P@5) and top 10 documents (P@10) are also reported.

\begin{table*}[t]
\centering
\caption{Collections Characteristics}
\begin{tabular}{|c|c|c|c|c|c|c|} \hline
ID & Lang. & Collection & Queries (title only) & \#docs  & \#qrels \\ \hline
FA & Persian & Hamshahri 1996-2002 & CLEF 2008-09, topics 551-650 & 166,774 & 9,625 \\ \hline
\end{tabular}
\label{tab:dataset}
\end{table*}

\subsection{Comparing Different Morphological Processing Methods}

In this section we aim at evaluating the proposed SS4MCT method. To this end we compare the proposed SS4MCT with a number of dictionary-based CLIR methods; the 5-gram truncation method (SPLIT) proposed in \cite{Monz:2005}, rule-based query expansion (RBQE) based on inflectional/derivation rules from Farazzin machine translator\footnote{http://www.faraazin.ir}, and the STeP1 stemmer~\cite{Shamsfard:2010} are the morphological processing approaches for the retrieval system. On the other hand, we run another set of experiments without applying any morphological processing method similar to the Persian state-of-the-art CLIR methods. Iterative translation disambiguation (ITD) \cite{Monz:2005}, joint cross-lingual topical relevance model (JCLTRLM) \cite{Ganguly:2012}, top-ranked translation (TOP-1), and the bi-gram coherence translation method (BiCTM), introduced in \cite{Dadashkarimi:2014} (assume $|\bar{c}_i|=0$), are the baselines without any morphological processing units. 
As shown in Table~\ref{table:data2} BiCTM outperforms all the baselines when there is no morphological processing unit. Although the improvement compared to JCLTRLM is not statistically significant, for simplicity we assume this model as a base of comparisons in the next set of experiments. In other words, we study the effect of the morphological processing units on the performance of BiCTM. As shown in Table~\ref{table:data2} the performance of the CLIR task degraded when we use the SPLIT approach. It is due to expanding the query with irrelevant tokens (e.g., \textit{normal/abnormal}). 
RBQE suffers from a similar problem to some extent; for example \textit{jat} is a valid suffix for \textit{sabzi} (=vegetable) in Persian whereas it is an invalid suffix for \textit{ketab} (=book). The results demonstrate that SS4MCT outperforms all the baselines in terms of MAP. This is due to a couple of reasons; first the ability of SS4MCT at finding infixes along with other affixes, particularly in irregular inflections and second its ability at deriving the likelihood/relevance of the rules in the collection/query.

\begin{table}	
	\centering
\caption{Comparing different methods in dictionary-based CLIR. Superscripts show that the MAP improvements over baselines are statistically significant.}
\label{table:data2}

\begin{tabular}{|c|c|c|c|c|c|c|c|c|c|c|c|}
	\cline{1-12}
	
	\multicolumn{6}{|c|}{without morphological processing} 		&\multicolumn{6}{|c|}{with morphological processing} 		 	\\
	\cline{1-12} 
	id	&		& MAP			& \%M	& P@5	& P@10	&id & 		&MAP				& \%M	& P@5	& P@10	\\ \hline
		& Mono		& 0.383			&	& 0.640	& 0.605	&  & Mono	&0.384				&	& 0.640	& 0.605	\\ \hdashline
	1	& TOP 1		& 0.213			& 55.6	& 0.348	& 0.346	&5 & SPLIT	&0.223				& 58.2	& 0.362	& 0.363	\\
	2	& ITD		& 0.238$^{1}$		& 62.0	& 0.404	& 0.38	&6 & STEM	&0.247				& 65.0	& 0.412	& 0.401	\\
	3	& JCLTRLM	& 0.2523 		& 65.70	&0.4000	&0.3910	&7 & RBQE	&0.245				& 63.8	& 0.380	& 0.389	\\
	4	& BiCTM		& \textbf{0.257}$^{12}$	& 67.0	& 0.406	& 0.406	&8 & SS4MCT	&\textbf{0.268}$^{4567}$	& 69.8  & 0.412 & 0.411	\\ \hline
\end{tabular}	
\end{table}
\vspace{-0.2cm}




\vspace{-0.2cm}
\section{Conclusion and Future Works}
\label{Conclusion}
\vspace{-0.2cm}
In this paper we proposed a new method for statistical stemming in morphologically complex texts. SS4MCT extracts a number of morphological rules based on edit distances of a large number of word pairs from a collection. 
Evaluating SS4MCT on a dictionary-based English-Persian CLIR task demonstrates its effectiveness. Considering adjacency of the characters in the rules and evaluating the method on informal text mining remained as future works.

\vspace{-0.3cm}
\section{Acknowledgement}
\vspace{-0.3cm}
The author would like to thank Razieh Rahimi and the anonymous reviewers for their helpful comments and feedback.
\vspace{-0.3cm}
\bibliographystyle{splncs03.bst}
\scriptsize{\bibliography{bibfile}}

\end{document}